\documentclass[reprint,
nofootinbib,
amsmath,amssymb,
aps,
prl,
]{revtex4-2}

\usepackage{aas_macros}
\usepackage[english]{babel}
\usepackage{amsmath,amssymb,bm,epsfig,color,graphicx}
\usepackage{booktabs}
\usepackage[mathscr]{eucal}
\usepackage{slashed}
\usepackage{cancel}
\usepackage{yhmath}
\usepackage[colorlinks=true,linkcolor=blue,citecolor=blue,urlcolor=blue]{hyperref}
\usepackage{hyperref}
\usepackage{tikz,xcolor}

\usepackage{amsmath}
\usepackage{slashed}
\usepackage{comment}
\usepackage{cancel}
\usepackage{multirow}
\usepackage{tikz-feynman}
\tikzfeynmanset{compat=1.1.0}

\newcommand{\bea}{\begin{array}}
\newcommand{\eea}{\end{array}}
\newcommand{\beq}{\begin{eqnarray}}
\newcommand{\eeq}{\end{eqnarray}}

\newcommand{\Tr}{\text{Tr}}

\begin{document}
\title{
Cold Darkogenesis: Dark Matter and Baryon Asymmetry in Light of the PTA Signal 
\\
}

\author{
Kohei Fujikura$^{1}$,\footnote{
E-mail address: kfujikura@g.ecc.u-tokyo.ac.jp} 
Sudhakantha Girmohanta$^{2,3}$,\footnote{
E-mail address: } 
Yuichiro Nakai$^{2,3}$\footnote{
E-mail address: ynakai@sjtu.edu.cn} 
and Zhihao Zhang$^{2,3}\footnote{
E-mail address: }$\\*[10pt]
$^1${\it \normalsize
Graduate School of Arts and Sciences, University of Tokyo, Komaba,\\ Meguro-ku, Tokyo 153-8902,
Japan} \\*[3pt]
$^2${\it \normalsize 
Tsung-Dao Lee Institute, Shanghai Jiao Tong University, \\ No.~1 Lisuo Road, Pudong New Area, Shanghai 201210, China} \\*[3pt]
$^3${\it \normalsize 
School of Physics and Astronomy, Shanghai Jiao Tong University, \\ 800 Dongchuan Road, Shanghai 200240, China} \\*[3pt]
}

\begin{abstract}

We build upon the intriguing possibility that the recently reported nano-Hz gravitational wave signal
by Pulsar Timing Array (PTA) experiments is sourced by a strong first-order phase transition from a nearly conformal dark sector.
The phase transition has to be strongly supercooled to explain the signal amplitude, while the critical temperature has to be in the $\cal{O}$(GeV) range,
as dictated by the peak frequency of the gravitational wave spectrum. However, the resulting strong supercooling exponentially dilutes away any pre-existing baryon asymmetry and dark matter, calling for a new paradigm of their productions. We then develop a mechanism of cold darkogenesis that generates a dark asymmetry during the phase transition
from the textured dark $SU(2)_{\rm D}$ Higgs field.
This dark asymmetry is transferred to the visible sector via neutron portal interactions,
resulting in the observed baryon asymmetry.
Furthermore, the mechanism naturally leads to the correct abundance of asymmetric dark matter,
with self-interaction of the scale that is of the right order to solve the diversity problem in galactic rotation curves. Collider searches for mono-jets and dark matter direct detection experiments
can dictate the viability of the model. 

\end{abstract}

\maketitle


\textit{\textbf{Introduction.---}} The observed baryon asymmetry of the Universe and the presence of cold dark matter (DM)
are two primary motivations for considering extensions of the otherwise remarkably successful Standard Model (SM).
A possible clue to the nature of this new physics
may come from the nano-Hz gravitational waves (GWs)
that have been recently detected by the pulsar timing array (PTA) collaborations~\cite{NANOGrav:2023gor,NANOGrav:2023hvm, EPTA:2023fyk,Reardon:2023gzh,Xu:2023wog}.
One of the potential sources of those GWs is the merger of supermassive binary black holes (SMBHB),
but this scenario faces the challenge of the so-called ``final parsec problem'',
which is the stalling of the SMBHB evolution at about a parsec separation
due to the lack of efficient dynamical friction~\cite{Begelman:1980vb}.
Several attempts have been made toward its resolution, however, no consensus has been achieved~\cite{Berczik:2006tz, Lodato:2009qd, Khan:2013wbx, surti2023central}. Alternatively, the PTA signal may originate from other new physics phenomena,
such as a first-order phase transition~\cite{Nakai:2020oit,Bringmann:2023opz,Athron:2023xlk} (for different new physics scenarios, consult Refs.~\cite{NANOGrav:2023hvm, Ellis:2023oxs}).
It has been recently argued that a phase transition scenario may fit the observed spectral shape and amplitude better than the canonical SMBHB merger~\cite{Ellis:2023oxs, NANOGrav:2023hvm}.

Ref.~\cite{Fujikura:2023lkn} has explained the PTA signal in terms of a supercooled phase transition
in a  nearly conformal dark sector (see also Refs.~\cite{Madge:2023dxc, Megias:2023kiy, Salvio:2023ynn, Salvio:2023blb}).
It showed that a strong supercooling is essential to realize the amplitude of the PTA signal.
However, the nano-Hz peak frequency suggests a phase transition temperature of $\sim {\mathcal O}$(1) GeV.
Therefore, any previously generated baryon asymmetry is exponentially diluted away during the phase transition.
A similar problem may occur for the DM energy density.
Therefore, it is natural to ask if baryon asymmetry can be produced during the phase transition.
In the present work, we will consider this possibility, where a dark number asymmetry is generated first
and then shared with the visible sector via a portal interaction.
During the phase transition epoch, the SM sphaleron processes have frozen out, hence,
the portal interaction needs to violate the baryon number, and not just the lepton number that could be reprocessed
in the leptogenesis scenario~\cite{Fukugita:1986hr}.
As we will demonstrate, the strong supercooling in the dark sector provides a natural setup for ``\textit{cold darkogenesis}''
to take place, which is an amalgamation of the notions of cold baryogenesis~\cite{Turok:1990in, Krauss:1999ng, Garcia-Bellido:1999xos, Konstandin:2011ds, Servant:2014bla} and darkogenesis~\cite{Shelton:2010ta}.

The central idea of the current paper is as follows.
The dark sector is governed by a conformal dynamics in the deep ultraviolet with a large number of colors $N$.
In addition to that, the conformal sector is coupled to dark $SU(N_{\rm H})$ and $SU(2)_{\rm D}$ fields.
The latter couples to chiral dark fermions $L_{\chi}, \chi$ with an anomalous global number symmetry $U(1)_{\rm D}$.
The field content is shown in Tab.~\ref{tab:dark_sector}. As the dark sector evolves to the low-energy scale,
$SU(N_{\rm H})$ undergoes confinement,
which generates a mass gap and breaks the conformal invariance spontaneously.
Thereafter, an $SU(2)_{\rm D}$ doublet Higgs $H_{\rm D}$, which couples considerably to the dilaton,
develops a vacuum expectation value (VEV) together with the dilaton,
breaking $SU(2)_{\rm D}$ spontaneously and generating massive dark fermions.
As the conformal phase transition is associated with strong supercooling,
$H_{\rm D}$ experiences a rapid change of mass in a cold empty Universe,
which results in a rapid amplification of long-wavelength modes of $H_{\rm D}$.
This process is far from thermal equilibrium and generates numerous $H_{\rm D}$ configurations with non-vanishing winding numbers. These configurations then relax to the vacuum state in the presence of $SU(2)_{\rm D}$ gauge fields,
either by altering the $H_{\rm D}$ winding number or the Chern-Simons (CS) number of $SU(2)_{\rm D}$ gauge fields.
When the CS number changes, dark fermions are anomalously produced.
If the dark sector has CP violations, a net dark number emerges.
This dark number is transferred to the visible sector via a neutron portal effective operator
that violates the dark number and SM baryon number.
As the Universe cools down, this operator becomes ineffective,
and the asymmetry is separately conserved in the dark and visible sectors.  

As $SU(N_{\rm H})$ confines, vector-like dark quarks $f,\bar{f}$ in Tab.~\ref{tab:dark_sector}
form massive dark baryons $p_{\rm D}$ and dark pions $\pi_{\rm D}$.
It is conceivable that this dark baryon sector inherits the dark number asymmetry
generated through interactions with $L_{\chi}$ and $\chi$.
Hence, this scenario yields an asymmetric $SU(N_{\rm H})$ composite DM state ($p_{\rm D}$).
As a byproduct of the DM being part of the confined sector with a $\sim$ GeV confinement scale,
it is strongly self-interacting, via $\pi_{\rm D}$ mediation,
and coincidentally has the order of self-interaction cross-section
that can be desirable for the diversity of galactic rotation curves together with the small-scale structure problems~\cite{Tulin:2017ara, Ren:2018jpt, Kribs:2016cew, Girmohanta:2022dog, Girmohanta:2022izb}. 

As the symmetric component of the DM annihilates to the dark pions, it is crucial that $\pi_{\rm D}$ decays to the visible sector before the onset of the Big Bang Nucleosynthesis (BBN). This necessitates the introduction of a portal operator connecting the dark and visible sectors. The same portal operator is constrained however by the direct detection experiments searching for ${\cal O}(1-10)$ GeV DM. This scenario of asymmetry sharing at the dark phase transition scale is highly restrictive and can be probed by mono-jet searches at the Large Hadron Collider (LHC) and DM direct detection experiments. 

\begin{table}
    \centering
    \resizebox{0.4\textwidth}{!}{
    \begin{tabular}{c|c|c|c}  
      Fields  & $\, SU(N_{\rm H}) \,$ & $\, SU(2)_{\rm D} \,$ & $\, U(1)_{\rm D} \,$ \\[1ex]
      \hline
      \hline 
      $H_{\rm D}$ & {\bf 1} & $\bf 2$ & 0 \\[1ex]
      $L_{\chi, i} \equiv \begin{pmatrix}
          \psi_{1,i} \\
          \psi_{2,i}
      \end{pmatrix}$  & {\bf 1} & $\bf 2$ & 1 \\[1ex]
      $\chi_{1,i} \, ,$ $\chi_{2,i}$ & {\bf 1} & {\bf 1} & $-1$ \\[1ex]
      $f_j$ & $\bf N_{\rm H}$ & {\bf 1} & $1/N_{\rm H}$ \\[1ex]
      $\bar{f}_j$ & $\bf \overline{N}_{\rm H}$ & {\bf 1} & $-1/N_{\rm H}$
      \end{tabular}
      }
    \caption{\footnotesize The dark sector particles and their representations.
    Here, $i=1,2,\cdots,N_{D_{\rm L}}$, $j=1,2,\cdots,N_{D_{\rm B}}$ denote the generational indices.
    The total dark number is $D=D_{\rm L} + D_{\rm B}$, where $L_{\chi}, \chi$ carry $D_{\rm L}$ number, while $f,\bar f$ carry $D_{\rm B}$.}
    \label{tab:dark_sector}
\end{table}

\textit{\textbf{PTA signal from dark phase transition.---}} Let us begin by describing the first-order confining phase transition
in a nearly conformal dark sector generating GWs detected by the PTAs.
The low-energy effective description can be expressed in terms of the effective potential of the dilaton $\varphi$,
which is the pseudo-Nambu-Goldstone boson of the broken scale invariance. In the UV,
we start with a conformal theory with a large number of colors $N$ coupled together with $SU(N_{\rm H})$ gauge fields.
Owing to the asymptotic freedom, the effect of the latter is negligible in the UV,
while the $SU(N_{\rm H})$ fields confine at a certain energy scale as we evolve towards the IR.
This confinement scale depends on the dilaton as the running of the effective coupling of $SU(N_{\rm H})$,
denoted as $g_{\rm H}$, from a UV scale $k$ to a lower scale $Q \lesssim \varphi$ gets a contribution from the CFT
which confines at $\varphi$ and does not contribute to the running below it~\cite{vonHarling:2017yew},
\begin{align}
    \frac{1}{g_{\rm H}^2(Q, \varphi)} &= - \frac{b_{\rm CFT}}{8 \pi^2} \ln \left( \frac{k}{\varphi}\right) - \frac{b_{\rm H}}{8 \pi^2} \ln \left( \frac{k}{Q} \right) \ .
    \label{Eq:running}
\end{align}
Here, the $\beta$-function coefficients are $b_{\rm CFT}=-\xi N$, with $\xi$ being a positive constant, $b_{\rm H} \equiv b_{\rm YM}+b_{f}$, $b_{\rm YM} = 11N_{\rm H}/3$, and $b_f = -2 N_{D_{\rm B}}/3$.
The confinement scale of $SU(N_{\rm H})$ is then given by
\begin{align}
    \Lambda_{\rm H} (\varphi) = k \left( \frac{\varphi}{k} \right)^{-{b_{\rm CFT}}/{b_{\rm H}}} = \Lambda_{\rm H,0} \left( \frac{\varphi}{\varphi_{\rm min}}\right)^n \ ,
\end{align}
where $n \equiv -b_{\rm CFT}/b_{\rm H}$, and $\Lambda_{\rm H,0}$ denotes the $SU(N_{\rm H})$ confinement scale at present.
The $SU(N_{\rm H})$ condensate provides the dilaton with the following effective potential~\cite{vonHarling:2017yew, Fujikura:2019oyi, Girmohanta:2024kyx}:
\begin{equation}
\label{Eq:radionPotential}
    V_{\rm eff}(\varphi)=
    \left\{
\begin{array}{ll}
V_0+\frac{\lambda_\varphi}{4}\varphi^4-\frac{b_{\rm H}}{\eta}\Lambda^4_{\rm H,0}\left(\frac{\varphi}{\varphi_{\rm min}}\right)^{4n} & ;~\varphi \geq \varphi_c \ , \\[2ex]
V_0+\frac{\lambda_\varphi}{4}\varphi^4-\frac{b_{\rm H}}{\eta}\gamma_c^4\varphi_c^4 & ;~\varphi<\varphi_c \ .
\end{array}
\right.
\end{equation}
Here, $V_0$ is chosen to make the minimum of the potential vanishing, the quartic term is scale-invariant
and is present in general, $\gamma_c \simeq \pi$, $\eta$ is determined from the numerical value of the $SU(N_{\rm H})$ condensate, 
$\varphi_c$ encodes the threshold between the confining and deconfining phases that can be determined
from the continuity of Eq.~\eqref{Eq:radionPotential}, and $\varphi_{\rm min}$ denotes the dilaton VEV.\footnote{
For further details, we refer the reader to Ref.~\cite{Fujikura:2019oyi}. Here we have taken Dirac $f$ to be at the cut-off and ignored the contribution of its condensate. If it is included, even more parameter space opens up~\cite{vonHarling:2017yew}.}
The confinement scale $\Lambda_{\rm H,0}$ in Eq.~\eqref{Eq:radionPotential} essentially sets the scale of the dilaton VEV,
\begin{align}
    \varphi_{\rm min} = \left( \frac{ 4 n b_{\rm H}}{\eta \lambda_\varphi} \right)^{1/4} \Lambda_{\rm H,0} \ .
    \label{Eq:phi_min}
\end{align}
The dilaton acquires a mass due to the stabilizing potential,\footnote{
A possible doubly composite dynamics may simultaneously address the gauge hierarchy problem
and result in a double-peaked GW spectrum~\cite{Girmohanta:2022giy}.}
\begin{align}
m_{\varphi}^2 & \equiv \frac{1}{Z} \frac{\partial^2 V_{\rm eff}(\varphi)}{\partial \varphi^2} \bigg |_{\varphi_{\rm min}} = \frac{8 \pi^2}{3 N^2}  (1-n) \lambda_{\varphi} \varphi_{\rm min}^2 \ ,
\label{Eq:mphi}
\end{align}
where a rescaling factor of $Z ={{3 N^2}/{2\pi^2}}$ ensures that the canonical kinetic term for the dilaton is properly normalized.

When the ambient temperature is higher than $\Lambda_{\rm H,0}$, the system is in the hot thermal deconfined phase.
As the temperature drops, the $SU(N_{\rm H})$ gauge field confines,
driving a first-order confinement-deconfinement phase transition in the dark sector.
The phase transition dynamics with the dilaton potential \eqref{Eq:radionPotential} has been analyzed in 
Ref.~\cite{Fujikura:2023lkn} which utilized the weakly coupled dual description due to 
AdS/CFT~\cite{Maldacena:1997re,Gubser:1998bc,Witten:1998qj}.
The collision of bubbles of the true vacuum and subsequent fluid flows produce shear stresses that source the GWs.
Here, we sketch the description of the key quantities that determine the GW spectra,
while we refer the reader to Ref.~\cite{Fujikura:2023lkn} for more details.
The critical temperature $T_c$ is defined as the temperature when the two phases have equal free energy,
while the nucleation temperature $T_n$ marks the moment
when the bubble nucleation rate of the true vacuum becomes equal to the Hubble expansion rate of the Universe.
The latter can be evaluated from the bounce action $S_{\rm B}$.
After the completion of the phase transition, the dark sector decays into the SM leaving behind only the DM,
as we will see later, and the Universe is reheated to the temperature,
\begin{align}
    T_{\rm RH} & \simeq \left( \frac{30}{\pi^2 g_*(T_{\rm RH})} \left|V_{\rm eff}(0)-V_{\rm eff}(\varphi_{\rm min})\right| \right)^{1/4} \ ,
    \label{TRH_eq}
\end{align}
where $g_*$ is the effective number of degrees of freedom.

\begin{figure}[!t]
    \centering
    \includegraphics[width=0.48\textwidth]{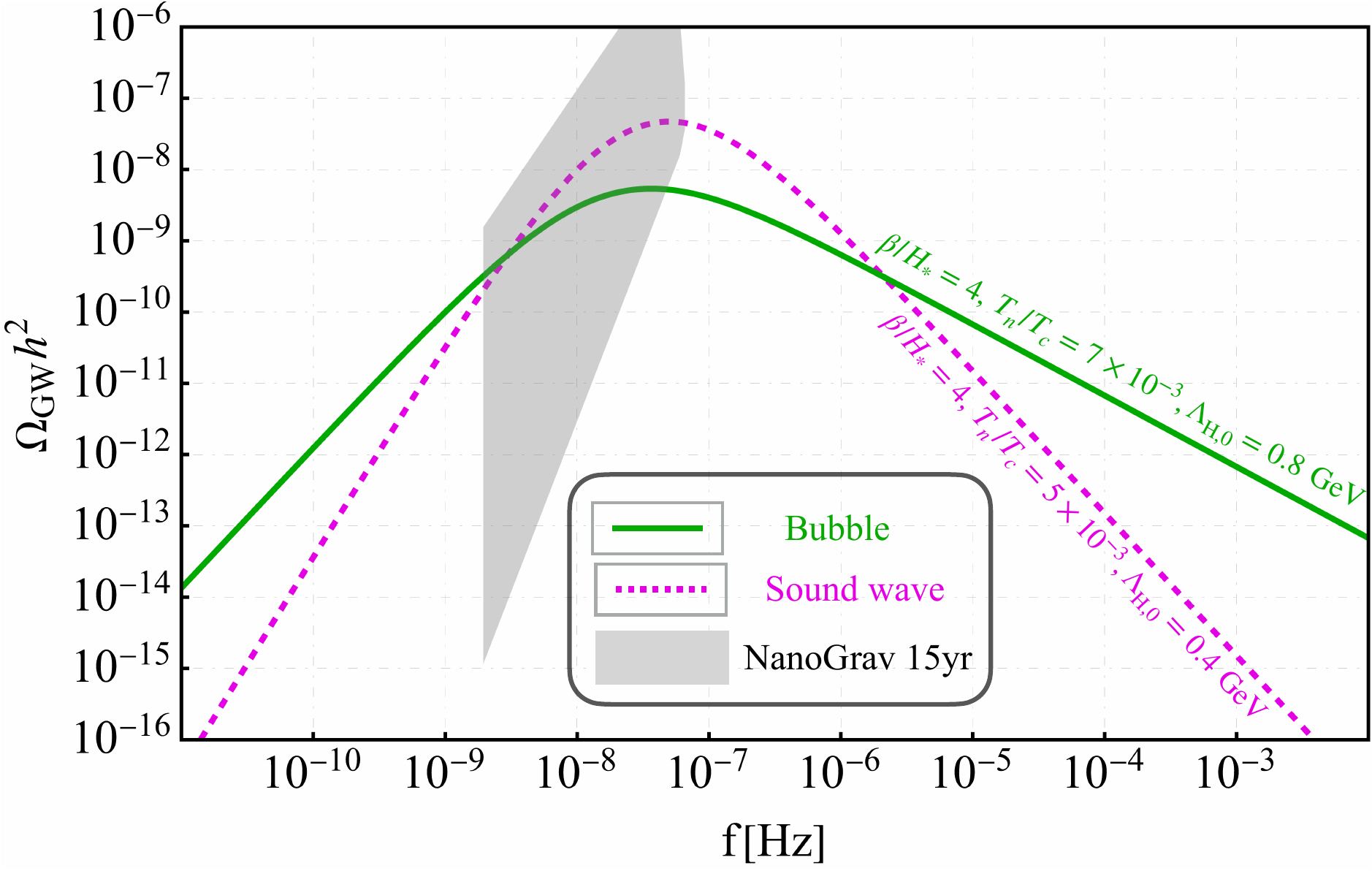}
    \caption{Spectra of a stochastic GW background produced by the dark conformal phase transition.
    The solid (dashed) curve represents the bubble collision (sound wave) only case.
    The gray-shaded region refers to the NANOGrav 15-yr signal region~\cite{NANOGrav:2023gor}.
    We have chosen $\lambda_\varphi=1$, $\eta=8$, $N=10$, $N_{\rm H}=5$, $N_{D_{\rm B}}=10$, $n=0.15$, $\Lambda_{\rm H,0}=0.8$ GeV, and $\lambda_\varphi=1$, $\eta=8$, $N=13$, $N_{\rm H}=6$, $N_{D_{\rm B}}=10$, $n=0.28$, $\Lambda_{\rm H,0}=0.4$ GeV for the bubble collision and sound wave only cases,
    respectively.}
    \label{fig:Grav_plot}
\end{figure}

Two key quantities acting as proxies for the phase transition dynamics are the latent heat released ($\alpha$)
and the inverse duration of the phase transition ($\beta$),
\begin{equation}
    \alpha \equiv \frac{V_{\rm eff}(0)-V_{\rm eff}(\varphi_{\rm min})}{\rho_{\rm rad}(T_n)} \ ; \quad \beta = H(T_n) \frac{d S_{\rm B}}{dT} \bigg |_{T_n} \ ,
\end{equation}
where $\rho_{\rm rad}$ symbolizes the energy density of the SM radiation bath.
Due to strong supercooling, $\alpha \gg 1$.
The GW spectra $\Omega_{\rm GW}$ are calculated by using the well-known approximate analytical estimates
summarized in the supplemental material.
In Fig.~\ref{fig:Grav_plot}, we show two representative fits to the GW spectra for the case
when the dominant contribution results from bubble collisions
or the sound wave in plasma.
The contribution due to the induced turbulence in plasma is negligible in comparison.
For both the bubble collision only and the sound wave only cases,
our parameter choices yield $\beta/H(T_{\rm RH}) \simeq 4$,
which is safe from possible overproduction of primordial black holes~\cite{Ellis:2023oxs}.
We can see that both cases well fit the NANOGrav 15-yr data~\cite{NANOGrav:2023gor}.

A crucial feature in explaining the PTA signal from a dark first-order phase transition is a considerable amount of supercooling,
as exemplified by the ratio of the nucleation and critical temperatures $T_n/T_c \sim 5 \times 10^{-3}$
in Fig.~\ref{fig:Grav_plot}.
Therefore, any pre-existing baryon asymmetry or DM energy density is diluted by a factor $\sim 10^{-7}$. We will provide a solution to this conundrum, where the baryon asymmetry and DM are created after the phase transition.
We assume the bubble wall velocity $v_{w} \sim 1$ for the GW estimate, while it is noted that frictions from particle splittings may lead to a terminal velocity for the wall~\cite{Bodeker:2017cim, Baldes:2020kam}. We will take the parameters used for the bubble collision only case as a benchmark in calculating the baryon asymmetry and DM production. Notice that with this parameter choice,
using Eqs.~\eqref{Eq:phi_min}-\eqref{TRH_eq},
we get $\varphi_{\rm min}=0.75$ GeV, $m_{\varphi}=0.35$ GeV, and $T_{\rm RH}=0.28$ GeV.

\textit{\textbf{Production of dark asymmetry.---}} 
The field content of the present model is summarized in Tab.~\ref{tab:dark_sector},
where we consider $SU(2)_{\rm D}$ gauge fields and chiral matter representations coupled to the CFT.
The dark Higgs field $H_{\rm D}$ has the following coupling with the dilaton~\cite{Konstandin:2011dr, Servant:2014bla}:
\begin{equation}
    V(\varphi, H_{\rm D}) =  V_{\rm eff}(\varphi) + \frac{\lambda}{4} \bigg[ H_{\rm D}^\dagger H_{\rm D} - \frac{v_{\rm D}^2}{2} \left( \frac{\varphi}{\varphi_{\rm min}} \right)^2\bigg]^2 \ .
    \label{Eq:HDphiPotential}
\end{equation}
Note that, when $\varphi$ is stuck at the false vacuum  $\varphi=0$, $H_{\rm D}$ does not develop a VEV.
Typically, this happens for a considerable amount of time, where the false vacuum energy stored in $\varphi$ essentially dominates the Universe, until $\varphi$ can tunnel to the true vacuum. We will assume a little hierarchy $v_{\rm D} \lesssim \varphi_{\rm min}$ such that one can neglect $H_{\rm D}$ contribution to the dilaton-driven phase transition. We now describe how a dark number asymmetry is generated during the phase transition.

Once $\varphi$ tunnels to the true minimum, suddenly, $H_{\rm D}$ experiences a spinodal instability,
and the dilaton energy is dumped into long wavelength modes of $H_{\rm D}$ that can carry a non-vanishing winding number~\cite{Konstandin:2011ds}.
Non-perturbative analysis reveals that the produced low momenta modes of Higgs and gauge bosons soon reach an effective thermal equilibrium by rescattering process with the effective temperature $T_{\rm D}$, which is higher than the reheating temperature~\cite{Khlebnikov:1996mc,Prokopec:1996rr}.
Following Refs.~\cite{Garcia-Bellido:1999xos, Konstandin:2011ds}, we estimate $T_{\rm D}$ from the energy density released through the spinodal instability, while utilizing the fact that $H_{\rm D}$ modes are populated till a momenta cut-off determined from Eq.~\eqref{Eq:HDphiPotential}
\begin{align}
    \frac{T_{\rm D}}{\varphi_{\rm min}}\sim 0.05\times \left(\frac{\varphi_{\rm min}}{\lambda v_{\rm D}}\right)^{1/2}.
    \label{Eq:TD}
\end{align}
%
Once $H_{\rm D}$ gets a VEV, due to the following Yukawa couplings, the dark leptons $\psi, \chi$ form massive Dirac states, 
\begin{align}
    -{\cal L}_{\rm Yuk} &= y_{ij}^{1} {H}_{\rm D} L_{\chi, i}  \chi_{1,j} + y_{ij}^{2} {H}_{\rm D}^{\dagger} L_{\chi, i}  \chi_{2,j} + {\rm h.c.} \ .
    \label{eq:LYuk}
\end{align}
The global $U(1)_{\rm D_L}$ for these leptons as noted in Tab.~\ref{tab:dark_sector} results into a chiral anomaly
for the corresponding global current $j_{\rm D_L}$ as follows,
\begin{equation}
    \partial_{\mu} j^{\mu}_{\rm D_L} = N_{\rm D_L} \frac{g_{\rm D}^2}{32 \pi^2} \Tr\left(W_{\rm D}^{\mu \nu} \widetilde{W}_{\rm D, \mu \nu}\right) \ .
    \label{Eq:chiralAnomaly}
\end{equation}
Here, $g_{\rm D}$, $W_{\rm D}^{\mu \nu}$ ($\widetilde{W}_{\rm D}^{\mu \nu}$) denote the $SU(2)_{\rm D}$ gauge coupling,
and (dual) field strength tensors respectively, while the number of generations for the dark leptons $N_{\rm D_L}$
is even to avoid a global $SU(2)_{\rm D}$ anomaly.
Eq.~\eqref{Eq:chiralAnomaly} is the source of the anomalous production of the dark leptons
in the presence of dynamics changing the gauge-Higgs winding number.

We introduce a CP-violating coupling with the gauge field which is of the form,
\begin{align}
	\mathcal{O}_{\rm CPV}= \delta_{\rm CP}\dfrac{H_{\rm D}^\dag H_{\rm D}}{\Lambda_{\rm CP}^2}\dfrac{g_{\rm D}^2}{32\pi^2} \Tr\left(W_{\rm D}^{\mu \nu} \widetilde{W}_{\rm D, \mu \nu}\right),
\label{eq:OCPV}
\end{align}
where $\delta_{\rm CP},~\Lambda_{\rm CP}$ respectively denote a dimensionless CP-violating phase and a new physics mass scale
which leads to the operator~\eqref{eq:OCPV} in the low-energy limit.
In the presence of this operator, the time-dependent Higgs condensation can be regarded as
the chemical potential of the CS number of the gauge field,
which results in the generation of a net non-zero dark lepton asymmetry through its chiral anomaly in Eq.~\eqref{Eq:chiralAnomaly}.
By solving the Boltzmann-like equation of the dark lepton number,
one can approximately estimate the produced dark lepton asymmetry in $L_{\chi}$, $\chi$ as~\cite{Garcia-Bellido:1999xos} (see Ref.~\cite{Tranberg:2006dg} for the non-perturbative numerical result)
\begin{align}
	{\cal D}_{\rm L, in} \simeq N_{D_{\rm L}}\dfrac{45\alpha_{\rm D}^4\delta_{\rm CP}}{\pi^2g_*} \dfrac{\langle H_{\rm D}^\dag H_{\rm D}\rangle}{\Lambda_{\rm CP}^2}\left(\dfrac{T_{\rm D}}{T_{\rm RH}}\right)^3.
 \label{eq:DLin}
\end{align}
Here, $\alpha_{\rm D} \equiv g_{\rm D}^2/(4\pi)$, while $\langle H_{\rm D}^\dag H_{\rm D}\rangle =  v_{\rm D}^2/2$.
Note that $\alpha_D^4$ factor comes from the dark $SU(2)_{\rm D}$ sphaleron transition rate.

The equivalent of three of the Sakharov conditions~\cite{Sakharov:1967dj} are satisfied as follows. $C$, $P$,
and the dark lepton number violation are due to the chiral representation of the dark sector and the chiral anomaly
as embodied in Eq.~\eqref{Eq:chiralAnomaly}. $CP$ violation originates from the operator~\eqref{eq:OCPV},
while the spinodal instability of $H_{\rm D}$ and the dynamics of gauge-Higgs winding configurations provide the out-of-thermal equilibrium condition. Combining the above equations, we have the following estimate:
\begin{multline}
{\cal D}_{\rm L, in}  \simeq  \  10^{-10}  \left( \frac{N_{D_{\rm L}}}{2} \right) \left(\frac{\delta_{\rm CP}}{10^{-4}} \frac{\varphi_{\rm min}^2}{\Lambda_{\rm CP}^2} \right) 
 \left( \frac{\alpha_{\rm D}}{1.5 \times 10^{-2}}\right)^4  \\ \times \left(\frac{\lambda}{10^{-4}}\right)^{-3/2}  \left(\frac{5 v_{\rm D}}{\varphi_{\rm min}} \right)^{1/2} \left( \frac{\varphi_{\rm min}}{3 T_{\rm RH}}\right)^3,
\label{Eq:DLin_numerical}
\end{multline}
where we have used $g_{*}=80$. This asymmetry produced in the dark sector will be shared with the visible sector via portal operators.

\textit{\textbf{Asymmetry sharing.---}} The generated asymmetry ${\cal D}_{\rm L, in}$ in the dark lepton sector $L_{\chi}$, $\chi$ in Eq.~\eqref{eq:DLin} is shared with the dark baryon sector via the effective operator,
\begin{equation}
    {\cal O}_{\rm D} \sim \frac{1}{\Lambda_{\rm D}^2} p_{\rm D} p_{\rm D} \chi \chi \ ,
    \label{Eq:ODD}
\end{equation}
where the effective scale $\Lambda_{\rm D}$ is expected to be at the GeV scale.
This is not contradictory from a UV completion perspective, as all fields are dark.
Further, the dark lepton asymmetry is shared with the visible sector via the neutron portal operator,
\begin{equation}
    {\cal O}_{n} \sim \frac{1}{\Lambda_n^2} \chi u_{\rm R} d_{\rm R} d_{\rm R} \ .
    \label{Eq:On}
\end{equation}
 To forbid bound neutron decays, we conservatively demand that $\chi$ is more massive than a neutron.\footnote{
 If higher generation SM quarks are used for the asymmetry sharing, this constraint on mass of $\chi$ can be relaxed.
 For related applications to the neutron lifetime anomaly and strange baryons, see Ref.~\cite{Alonso-Alvarez:2021oaj}.}
 $\Lambda_{n} \lesssim 15$ TeV is adequate to keep the operator in Eq.~\eqref{Eq:On} in thermal equilibrium at the GeV temperature scale, while also allowing $\psi, \chi$ to decay before the onset of the BBN. 

The operator ${\cal O}_{n}$ violates the baryon number $B$ and the total dark number $D=D_{\rm L}+ D_{\rm B}$ separately,
but conserves the lepton number $L$, and the generalized baryon number $B+D$.
The global symmetry $U(1)_{\rm B+D}$ has to be preserved with high-quality
to prevent a Majorana mass term $\chi \chi$ that could cause the washout of the asymmetry.
In terms of the dual description,
this is realized as a 5D bulk gauge symmetry $U(1)_{\rm B+D}$ broken on the UV brane.
Further, a discrete symmetry $\mathbb{Z}_2^{\rm D}$, under which only $p_{\rm D}$ is odd, is sufficient to forbid terms such as $p_{\rm D} \chi^n$, where $n=1,3$,
etc., which could cause the DM to decay.

We can utilize the fact that the operators ${\cal O}_{\rm D}$ and ${\cal O}_{n}$ were in thermal equilibrium
at one point of the evolution together with the condition of the conservation of $B+D$,
modulo the $SU(2)_{\rm D}$ anomaly that generates the dark lepton asymmetry ${\cal D}_{\rm L, in}$,
and electromagnetic charge neutrality to estimate the shared asymmetry in the visible sector (${\cal B}_f$)
and the dark baryon sector (${\cal D}_{\rm B}$) as
\begin{align}
     \nonumber
    {\cal B}_f &= \left[\frac{2+4 N_{D_{\rm L}}}{4 N_{D_{\rm L}} + N_{D_{\rm B}}+2} \right] {\cal D}_{\rm L, in} \ ,  
  \\[1ex]
   {\cal D}_{\rm B} &= \left[\frac{N_{D_{\rm B}}}{4 N_{D_{\rm L}} + N_{D_{\rm B}}+2} \right] {\cal D}_{\rm L, in}  \ .
   \label{Eq:asymmResult}
\end{align}
The details of this derivation are presented in the supplemental material.
Notice that for the asymmetric DM to make up the observed energy density of DM, we should have
\begin{align}
    m_{p_{\rm D}} \simeq 5 \bigg| \frac{{\cal B}_f}{{\cal D}_{\rm B}} \bigg| {\rm GeV} = 5 \left(\frac{2+4 N_{D_{\rm L}}}{ N_{D_{\rm B}}} \right) {\rm GeV} \ .
\end{align}


\begin{figure}[t]
    \centering
    \includegraphics[width=0.48\textwidth]{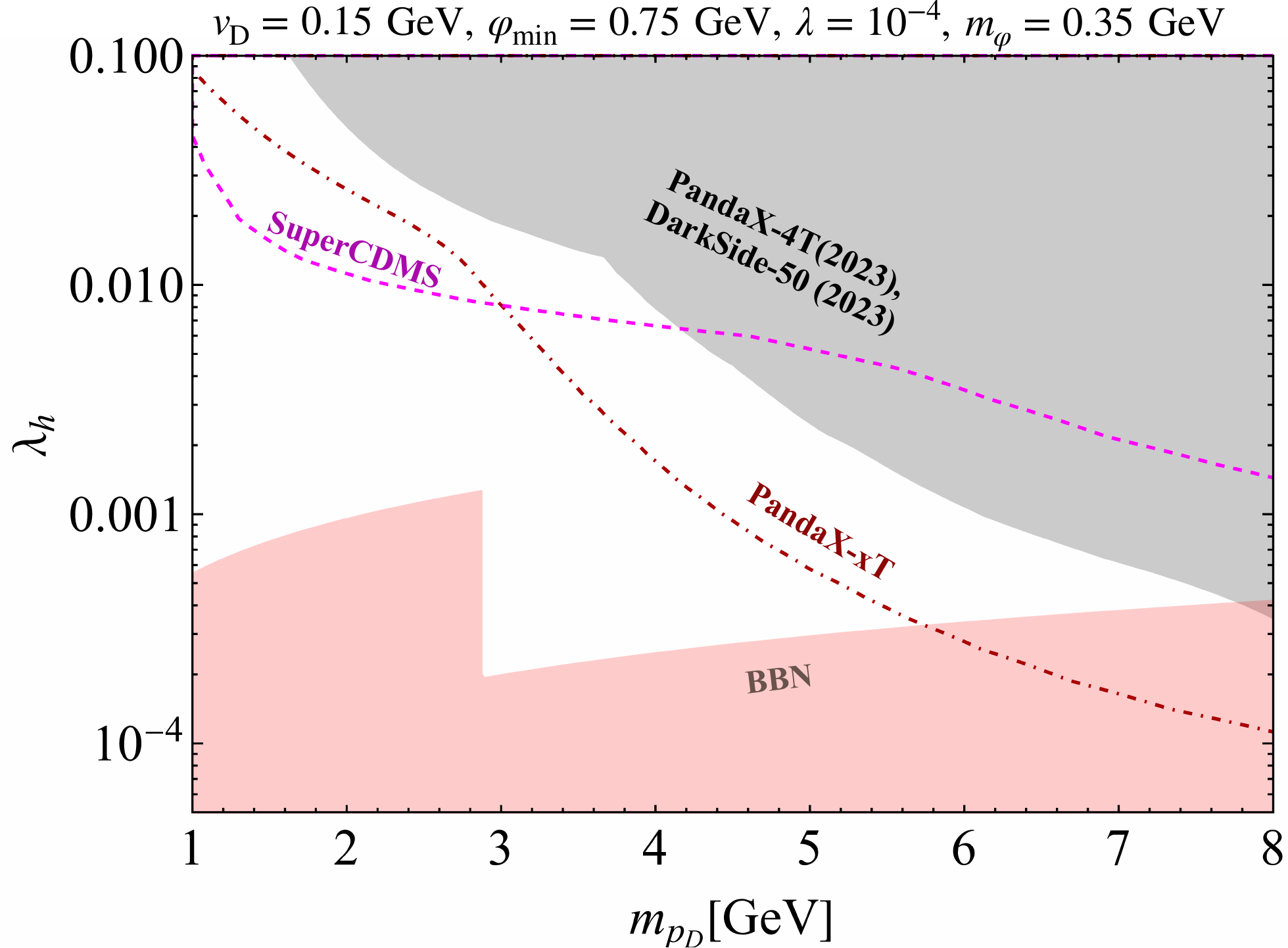}
    \caption{\footnotesize DM direct detection constraint (gray) and BBN constraint (red) on the Higgs portal coupling in Eq.~\eqref{Eq:HiggsPortal} as a function of the DM mass.
    Here, $N_{\rm H}=5$, $\varphi_{\rm min} = 0.75$ GeV and $m_{\varphi}=0.35$ GeV are input from the bubble collision only case
    in Fig.~\ref{fig:Grav_plot}, while we take $v_{\rm D}=0.2 \varphi_{\rm min}$ and $\lambda=10^{-4}$
    that result in $T_{\rm D}=8$ GeV.
    The magenta dashed and red dot-dashed curves correspond to the projected reach from
    SuperCDMS~\cite{Zatschler:2024ssq} and PandaX-xT~\cite{PandaX:2024oxq}.}
    \label{fig:pltNG}
\end{figure}

\textit{\textbf{Phenomenology.---}} As the DM is a composite state in an $SU(N_{\rm H})$ gauge theory,
it can interact with itself via the mediation of dark pions. Let us denote the effective scattering length as $a_{\rm D}$.
The self-interaction cross-section can therefore be estimated as~\cite{Tulin:2017ara, Kribs:2016cew}
\begin{align}
    \frac{\sigma_{p_{\rm D} p_{\rm D}}}{m_{p_{\rm D}}} \sim 1 \  {\rm cm}^2/{\rm g} \left( \frac{\Lambda_{\rm H,0}}{m_{p_{\rm D}}} \right) \left( \frac{\Lambda_{\rm H, 0}}{a_{\rm D}^{-1}}\right)^2 \left( \frac{150 \ {\rm MeV}}{\Lambda_{\rm H,0}} \right)^3 \ ,
\end{align}
which is in the ballpark value as desired from the small-scale structure issues,
such as the observed diversity in the galactic rotation curves.
Furthermore, as the constituent dark quark is Dirac, the cross-section should also fall off beyond the inverse momentum scale
$\sqrt{\sigma_{p_{\rm D} p_{\rm D}}}$ as required to be consistent with the observations at the cluster scale.
In particular, from Ref.~\cite{Cline:2013zca}, $\Lambda_{\rm H,0} < 4$ GeV and $m_{p_{\rm D}} < 15$ GeV is required
to have this property, which is satisfied here.
Note that these naive scaling arguments may break due to the strongly interacting nature of the constituents,
and we leave the detailed analysis of DM self-interactions in the current model as a future study.\footnote{See Ref.~\cite{Cline:2013zca} for a discussion on the self-interaction strength in dark confining theories from lattice gauge studies.} 

Let us comment on the searches for the portal operators.
Note that $\Lambda_{n}$ in Eq.~\eqref{Eq:On} can be probed at colliders in the partonic processes
$ud \to \bar \chi \bar d$, $dd \to \bar u \bar \chi$ and their resulting mono-jet signatures.
The existing bound already constrains $\Lambda_n \gtrsim 2$ TeV~\cite{ATLAS:2021kxv, Ciscar-Monsalvatje:2023zkk} and
will be constrained further by the High-Luminosity LHC.
As the symmetric DM density ends up annihilating into the dark pions, via the process
$p_{\rm D} + \bar p_{\rm D} \to \pi_{\rm D} \pi_{\rm D}$, they should decay to the visible sector before the onset of the BBN.
This entails the introduction of a portal operator connecting the dark and visible sectors. We take the Higgs portal operator,
\begin{equation}
    {\cal L}_{\rm H} \supset -\lambda_{h} \left(|H|^2-\frac{v^2}{2}\right) \left(|H_{\rm D}|^2 -\frac{v_{\rm D}^2}{2} \frac{\varphi^2}{\varphi_{\rm min}^2} \right) \ . 
    \label{Eq:HiggsPortal}
\end{equation}
To be consistent with the constraint on Higgs invisible decays, $\lambda_h \lesssim 0.1$~\cite{ATLAS:2023tkt}.
The DM direct detection places a more stringent upper limit on this coupling, while the BBN constraint sets a lower bound.
This is depicted in Fig.~\ref{fig:pltNG} for an illustrating set of parameters, $v_{\rm D}=0.2 \varphi_{\rm min}$ and $\lambda=10^{-4}$ together with $\varphi_{\rm min} = 0.75$ GeV, $N_{\rm H}=5$ and $m_{\varphi}=0.35$ GeV,
as taken from the bubble collision only case in Fig.~\ref{fig:Grav_plot}.
The gray region is excluded by PandaX-4T~\cite{PandaX:2022aac} and Darkside-50~\cite{DarkSide-50:2022qzh} experiments,
and the red region is constrained from the BBN.
The kink in the BBN region appears once the dark pion can decay to charm final states.
The current parameter choices yield $T_{\rm D} \simeq 8$ GeV,
allowing for the production of $p_{\rm D}$ for the entirety of the allowed parameter space in Fig.~\ref{fig:pltNG}.
For example, $N_{D_{\rm L}}=2$, $N_{D_{\rm B}}=10$ yields the observed baryon asymmetry in Eq.~\eqref{Eq:asymmResult}
with the parameter choices made in Figs.~\ref{fig:Grav_plot},~\ref{fig:pltNG} and a $5$ GeV DM candidate.
We also depict the projected reach from SuperCDMS~\cite{Zatschler:2024ssq} and PandaX-xT~\cite{PandaX:2024oxq},
which will probe the parameter space further.
For details of the calculation and parameter dependence, refer to the supplemental material.

In summary, we have presented a concrete scenario of producing baryon asymmetry and DM
in a dark supercooled phase transition that can explain the PTA signal.
The baryon asymmetry is produced after the phase transition, which also yields an asymmetric self-interacting DM.
Our model will be probed by near-future DM direct detection experiments and colliders.

\textit{\textbf{Acknowledgements.---}} We are grateful to Gabriel Cardoso and Jianglai Liu for useful discussions.
KF is supported by JSPS Grant-in-Aid for Research
Fellows Grant No.\,22J00345.

\bibliographystyle{apsrev4-2}
\bibliography{bib}

\end{document}


\begin{CJK*}{UTF8}{gbsn} 
\preprint{APS/123-QED}

\title{Supplemental Material:\\
Cold Darkogenesis: Dark Matter and Baryon Asymmetry in Light of the PTA Signal}

\author{
Kohei Fujikura$^{1}$,\footnote{
E-mail address: kfujikura@g.ecc.u-tokyo.ac.jp} 
Sudhakantha Girmohanta$^{2,3}$,\footnote{
E-mail address: } 
Yuichiro Nakai$^{2,3}$\footnote{
E-mail address: ynakai@sjtu.edu.cn} 
and Zhihao Zhang$^{2,3}\footnote{
E-mail address: }$\\*[10pt]
$^1${\it \normalsize
Graduate School of Arts and Sciences, University of Tokyo, Komaba,\\ Meguro-ku, Tokyo 153-8902,
Japan} \\*[3pt]
$^2${\it \normalsize 
Tsung-Dao Lee Institute, Shanghai Jiao Tong University, \\ No.~1 Lisuo Road, Pudong New Area, Shanghai 201210, China} \\*[3pt]
$^3${\it \normalsize 
School of Physics and Astronomy, Shanghai Jiao Tong University, \\ 800 Dongchuan Road, Shanghai 200240, China} \\*[3pt]
}



\maketitle
\end{CJK*}

\subsection{Calculation of the GW spectra}

A first-order phase transition proceeds via the nucleation of bubbles,
which subsequently expand due to the pressure difference between the false and true vacua.
Essentially, the collision of bubbles or the sound wave generated in the plasma source the shear stresses
that produce GWs, denoted as $\Omega_{\rm coll}$ and $\Omega_{\rm sound}$, respectively.
A possible subdominant contribution comes from the turbulence of the plasma, which we neglect.
These GW spectra are evaluated from the latent heat released $\alpha$ and the inverse time duration $\beta$
as defined in the main text.
The contribution from bubble collisions is given by~\cite{Jinno:2016vai, NANOGrav:2023hvm}
\begin{align}
\nonumber
& \Omega_{\rm coll} h^2 \simeq 5.5 \times 10^{-7} \left(\dfrac{20}{g_*(T_{\rm RH})}\right)^{\frac{1}{3}} \left(\frac{H(T_{\rm RH})}{\beta} \right)^2   \\  & \qquad \qquad \times \left( \frac{\alpha}{1+\alpha}\right)^2\mathcal{S}(f/f_{\rm coll}) \ ,\nonumber
\\[1ex]
& \mathcal{S}(x)= \dfrac{1}{\mathcal{N}}\dfrac{(a+b)^c}{\big[b x^{-a/c} + a x^{b/c}\big]^c} \ ,\label{bubble:eq}\\[1ex]
& \mathcal{N} =  \left(\dfrac{b}{a}\right)^{a/d} \left(\dfrac{dc}{b}\right)^c \dfrac{\Gamma(a/d)\Gamma(b/d)}{\Gamma(c) d}, \quad d=\dfrac{a+b}{c} \ .\nonumber
\end{align}
Here $\Gamma$ is the Euler Gamma function, $g_*$, $H$ are the effective number of degrees of freedom
and the Hubble rate at the reheat temperature $T_{\rm RH}$, respectively,
and $(a, b, c)$ denote the bubble spectral shape parameters.
From the maximum posterior values in NANOGrav analysis~\cite{NANOGrav:2023hvm},
we take $(a,b,c)=(1.97, 1, 3 )$. The sound wave contribution is given as follows:
\begin{align}
\nonumber
& \Omega_{\rm sound}  h^2 \simeq 1.4\times 10^{-6}v_w\left(\dfrac{20}{g_*(T_{\rm RH})}\right)^{\frac{1}{3}}\left(\frac{H(T_{\rm RH})}{\beta} \right) \\ &  \times \left( \frac{\kappa_{\rm sw} \alpha}{1+\alpha}\right)^2\left(1-\frac{1}{\sqrt{1+2 t_{\rm sw} H(T_{\rm RH})}} \right)\mathcal{S}(f/f_{\rm sw})  \ .
\label{sw:eq}
\end{align}
We take maximum posterior values, $(a,b,c)= (3, 2, 5)$~\cite{NANOGrav:2023hvm}.
In Eqs.~\eqref{bubble:eq},~\eqref{sw:eq}, the peak frequencies $f_{\rm coll,sw}$ are defined as
\begin{align}
\nonumber
&f_{\rm coll,sw}\simeq 126\,{\rm nHz}\,
    \frac{1}{v_w}\left(\dfrac{g_*(T_{\rm RH})}{20}\right)^{\frac{1}{6}}\left(\dfrac{T_{\rm RH}}{1\,{\rm GeV}}\right) \\ 
  & {\hspace{1.3 cm}}  \times \dfrac{\beta}{H(T_{\rm RH})}f^{*}_{\rm coll,sw}\,,\\[1ex]
    &f^*_{\rm coll}\simeq 0.2 , \quad f^*_{\rm sw}\simeq 0.54 \ ,  \\[1ex]
  & t_{\rm sw}  \sim (8 \pi)^{1/3} \frac{v_w}{\beta \sqrt{3\kappa_{\rm sw} \alpha/4(1+\alpha) }} \ ,
  \label{GW_amp:eqs}
\end{align} 
where the suppression factor of the short period of the sound wave is included,
while $v_w$, $\kappa_{\rm sw}$ denote the bubble wall velocity and the efficiency factor, respectively.
Since the parameter space we are interested in results in a very large supercooling,
We assume $v_{w} \sim 1$ in natural units for the GW spectrum estimation.
In this limit, the efficiency factor $\kappa_{\rm sw}$ can be expressed as
\begin{align}
\kappa_{\rm sw} = \dfrac{\alpha}{0.73+0.083\sqrt{\alpha}+\alpha} \ .
\end{align}
%
\subsection{Details of asymmetry sharing calculation}

To calculate the asymmetries shared between the visible and dark sectors,
we follow the methods described in Ref.~\cite{Harvey:1990qw}.
We will use the fact that all the perturbative interactions in the model preserve the lepton number $L$
and the generalized baryon number $B+D_{\rm B}+D_{\rm L}$,
where $D_{\rm B}$ and $D_{\rm L}$ are the dark baryon and lepton numbers respectively as defined in the main text.
Only non-perturbative effects of the $H_{\rm D}$ configurations violate $D_{\rm L}$
and act as the source of the asymmetry that later gets reprocessed into the visible and dark baryon asymmetries
via the portal operators.
In the following, we will use the calligraphic notation ${\cal X}$ to represent the asymmetry in a generic quantum number $X$.

One starts with the observation that the number asymmetry of a particle species $\zeta$
can be traded off for its chemical potential $\mu_\zeta$ at temperature $T$ as
\begin{equation}
    \frac{n_{\zeta}-n_{\bar \zeta}}{s} = \frac{15 g_\zeta}{2 \pi^2 g_{*S}} \left(\frac{\mu_\zeta}{T}\right) \begin{cases}
        1 &   {(\rm boson)}\\[0.5ex]
        $1/2$ &  {(\rm fermion)}
        \end{cases} \ ,
        \label{Eq:asymmetry}
\end{equation}
where $g_\zeta$ denotes the number of internal degrees of freedom (d.o.f) for $\zeta$,
and $g_{*S}$ counts the total number of relativistic d.o.f. in the entropy density $s$.
After the breaking of $SU(2)_{\rm D}$, $H_{\rm D}$ chemical potential vanishes.
The Yukawa interaction $H_{\rm D}^{(\dagger)} L_\chi \chi_{1,2}$ then brings
the chemical potential of $\psi_{1,2}$ and $\chi_{1,2}$ to equality as a result.
Furthermore, the self-interaction of the $SU(2)_{\rm D}$ gauge bosons establishes chemical equilibrium among themselves.
Here, for simplicity, we will assume that the different generations share the same chemical potential.
The dark lepton and dark baryon asymmetries, modulo some constant factors,
can now be defined in terms of the corresponding chemical potentials as
\begin{align}
\nonumber
     {\cal D}_{\rm L} & = \sum_{i} \left( \mu_{\psi_{1,i}} + \mu_{\psi_{2,i}} - \mu_{\chi_{1,i}} - \mu_{\chi_{2,i}} \right)  \nonumber
    \\ &= -4  N_{D_{\rm L}} \mu_{\chi}  \ , \\[1.5ex]  
    {\cal D}_{\rm B} & = \sum_{j} \mu_{p_{{\rm D}, j}} = N_{D_{\rm B}} \mu_{p_{\rm D}} \ ,
    \label{Eq:DLDB}
\end{align}
where $N_{D_{\rm L}}$, $N_{D_{\rm B}}$ denote the numbers of dark lepton and dark baryon generations, respectively. 

We will only consider the SM particles that are ultra-relativistic at the reheat temperature $T_{\rm RH} \sim {\cal O} (1)$ GeV. The $u$, $d$, $s$ quarks and $e$, $\mu$ and neutrinos are ultra relativistic.
Note that for the production of the DM, the relevant temperature is the dark Higgs effective temperature $T_{\rm D}$,
which is $\sim 8$ GeV (see the main text).
Since the weak interactions freeze out below the MeV temperature scale,
we keep track of the chemical potential in the $W$ bosons that will be transferred to the quarks and leptons
after they decay.
We will soon see that due to the electromagnetic charge neutrality,
this chemical potential is zero.
As explained in the main text, a non-zero ${\cal D}_{\rm L}$ is generated from the $U(1)_{\rm D}$ anomaly
and acts as the initial condition in our following argument,
\begin{align}
	{\cal D}_{\rm L, in} \simeq N_{D_{\rm L}}\dfrac{45\alpha_{\rm D}^4\delta_{\rm CP}}{\pi^2g_*} \dfrac{\langle H_{\rm D}^\dag H_{\rm D}\rangle}{\Lambda_{\rm CP}^2}\left(\dfrac{T_{\rm D}}{T_{\rm RH}}\right)^3.
 \label{Eq:DLin}
\end{align}
The total lepton asymmetry (not to be confused with the Lagrangian) is given by
\begin{align}
    \nonumber
    {\cal L} & = \sum_{i} \left( \mu_{\nu_{\rm L}} + \mu_{e_{\rm L}} + \mu_{e_{\rm R}}\right) = 3 \mu_{\nu_{\rm L}} + 4 \mu_{e_{\rm L}} \\ 
    &= 7 \mu_{\nu_{\rm L}} + 4 \mu_{W} = 0 \ ,
    \label{Eq:lepton_asymmetry}
\end{align}
where we have used the fact that after the SM Higgs gets a VEV its chemical potential vanishes,
resulting in the equality of the left and right-handed chemical potentials for massive SM fields.
Furthermore, we have used that the weak interactions 
imply that
\begin{equation}
   \mu_W = \mu_{e_{\rm L}}-\mu_{\rm \nu_L} = \mu_{d_{\rm L}}-\mu_{u_{\rm L}} \ . 
   \label{Eq:muW}
\end{equation}
Now, as the $U(1)_{\rm em}$ in the SM is unbroken, the Universe has to be electromagnetically neutral.
The charge asymmetry is
\begin{align}
\nonumber
    {\cal Q}_{\rm em} &= \sum_{i} g_{\rm color} \bigg[ \frac{2}{3} \left( \mu_{u_{\rm L}}+ \mu_{u_{\rm R}} \right) -\frac{1}{3} \left( \mu_{d_{\rm L}}+ \mu_{d_{\rm R}} \right) \bigg]\nonumber
    \\
    & - \left(\mu_{e_{\rm L}} + \mu_{e_{\rm R}}\right) - 6 \mu_{W} = \frac{41}{2} \mu_{\nu_L} = 0 \ ,
    \label{Eq:charge_asymmetry}
\end{align}
where $g_{\rm color}=3$ and Eqs.~\eqref{Eq:lepton_asymmetry},~\eqref{Eq:muW} are used.
Therefore, together with the lepton number conservation and charge neutrality, we arrive at the condition, $\mu_{\nu_L} = \mu_{W}=0$.

Let us now consider the asymmetry-sharing operators that are in equilibrium, namely
\begin{align}
    \nonumber
    & {\cal O}_{\rm DD} \sim \frac{1}{\Lambda_{\rm D}^2} p_{\rm D} p_{\rm D} \chi \chi \ , \\[1ex]
    & {\cal O}_{n} \sim \frac{1}{\Lambda_n^2} \chi u_{\rm R} d_{\rm R} d_{\rm R}  \ .
    \label{Eq:sharingOps}
\end{align}
When they are in chemical equilibrium,
\begin{align}
\nonumber
   & \mu_{p_{\rm D}} + \mu_\chi = 0 \ , \\[1ex]
   & \mu_\chi = -\left(\mu_{u_{\rm R}} + 2 \mu_{d_{\rm R}}\right) = -3 \mu_{u_{\rm L}} \ .
   \label{Eq:chemicalSharing}
\end{align}
Hence, the visible and dark baryon asymmetries are 
\begin{align}
    \nonumber
  &  {\cal B} = \sum_{i} \mu_{u_{{\rm L}, i}} + \mu_{d_{{\rm L}, i}} + \mu_{u_{{\rm R}, i}} + \mu_{d_{{\rm R}, i}} = 6 \mu_{u_{\rm L}} \ , \\
  & {\cal D}_{\rm B} = N_{D_{\rm B}} \mu_{p_{\rm D}} =  3 N_{D_{\rm B}} \mu_{u_{\rm L}} \ , \nonumber 
  \\[1.5ex]
  & {\cal D}_{\rm L} = -4 N_{D_{\rm L}} \mu_\chi = 12 N_{D_{\rm L}} \mu_{u_{\rm L}} \ .
    \label{Eq:allAsymm}
\end{align}
These are the final resultant asymmetries after the chemical equilibrium persists for a finite amount of time
and the asymmetries are frozen out.
As the ${\cal B}+{\cal D}_{\rm B}+{\cal D}_{\rm L}$ is conserved by the perturbative fermion number violating processes
we consider, this quantity must be equal to the initial ${\cal D}_{\rm L, in}$ given in Eq.~\eqref{Eq:DLin}.
Then, we can now express all the final asymmetries in terms of the given input ${\cal D}_{\rm L, in}$ as
\begin{align}
     \nonumber
    {\cal B} &= \left[\frac{2}{4 N_{D_{\rm L}} + N_{D_{\rm B}}+2} \right] {\cal D}_{\rm L, in} \ , \nonumber 
  \\[1.5ex]
   {\cal D}_{\rm B} &= \left[\frac{N_{D_{\rm B}}}{4 N_{D_{\rm L}} + N_{D_{\rm B}}+2} \right] {\cal D}_{\rm L, in} \ , \nonumber
  \\[1.5ex]
   {\cal D}_{\rm L} &= \left[ \frac{4 N_{D_{\rm L}}}{ 4 N_{D_{\rm L}} + N_{D_{\rm B}}+2} \right] {\cal D}_{\rm L, in} \ .
   \label{Eq:asymmResult}
\end{align}
 The asymmetry stored in the dark lepton $\chi$, i.e., ${\cal D}_{\rm L}$, is transferred to the ordinary baryons as it decays. Therefore, the final baryon asymmetry is obtained as
 \begin{align}
     {\cal B}_{f} &= {\cal B} + {\cal D}_{\rm L} = \left[\frac{2+4 N_{D_{\rm L}}}{4 N_{D_{\rm L}} + N_{D_{\rm B}}+2} \right] {\cal D}_{\rm L, in} \ .
 \end{align}

\subsection{Details of phenomenological consequences}

The model contains three scalars that define the flavor basis, $\Phi = \left( h \ h_{\rm D} \ \tilde{\varphi} \right)^{\rm T}$, where $\tilde{\varphi} = \sqrt{Z} \varphi$ and $Z={3 N^2}/{2 \pi^2}$ ensures the canonical kinetic term for the dilaton.
They mix due to the following terms in the potential,
\begin{align}
    \nonumber
    V_{\Phi} & = V_{\rm eff}(\varphi) + \frac{\lambda}{4} \bigg[ H_{\rm D}^\dagger H_{\rm D} - \frac{v_{\rm D}^2}{2} \left( \frac{\varphi}{\varphi_{\rm min}} \right)^2\bigg]^2 \\
    & + \lambda_{h} \left(|H|^2-\frac{v^2}{2}\right) \left(|H_{\rm D}|^2 -\frac{v_{\rm D}^2}{2} \frac{\varphi^2}{\varphi_{\rm min}^2} \right) + \frac{1}{2} m_h^2 h^2 \ .
\label{Eq:Vtot_scalars}
\end{align}
The relevant temperature scale of the dark phase transition is $\sim 1$ GeV, hence,
the electroweak symmetry breaking has already taken place,
and we include the SM Higgs effective potential as a resulting mass term in Eq.~\eqref{Eq:Vtot_scalars}.
In the unitary gauge, the fluctuations $\Phi$ are defined around the minima of each field,
\begin{align}
\langle H \rangle= \frac{1}{\sqrt{2}} \begin{pmatrix}
                              0  \\
                              v
                      \end{pmatrix} \  ; \quad  \langle H_{\rm D} \rangle= \frac{1}{\sqrt{2}} \begin{pmatrix}
                              0  \\
                              v_{\rm D}
                      \end{pmatrix} \ ; \quad \langle \varphi \rangle = \varphi_{\rm min} \ .
\label{Eq:Minima_Phi}
\end{align}
Expanding Eq.~\eqref{Eq:Vtot_scalars} around the minima, one arrives at the potential in the flavor basis with the mass matrix, 
\begin{equation}
    V_{\Phi} = \frac{1}{2} \Phi^{\rm T} \cdot {\cal M} \cdot \Phi \ ; \quad {\cal M}_{ij} \equiv \frac{\partial^2 V_{\Phi}}{\partial \Phi_i \partial \Phi_j} \bigg |_{\langle \Phi \rangle} \ .
    \label{Eq:M_def}
\end{equation}
It can be explicitly checked that $\frac{\partial V_{\Phi}}{\partial \Phi_i}\big |_{\langle \Phi \rangle} = 0$, which is consistent with the assumption that Eq.~\eqref{Eq:Minima_Phi} represents the minima. The mass matrix in the flavor basis reads
\begin{align}
    {\cal M} & = \renewcommand{\arraystretch}{2}\begin{pmatrix}
        m_h^2 &  \lambda_h v v_{\rm D} & -\frac{\lambda_h v v_{\rm D}^2}{\sqrt{Z}\varphi_{\rm min}} \\[1mm]
    \lambda_h v v_{\rm D} & \frac{v_{\rm D}^2 \lambda}{2} & -\frac{v_{\rm D}^3 \lambda}{2 \sqrt{Z} \varphi_{\rm min}} \\[1mm]
    -\frac{\lambda_h v v_{\rm D}^2}{\sqrt{Z} \varphi_{\rm min}} & -\frac{v_{\rm D}^3 \lambda}{2 \sqrt{Z} \varphi_{\rm min}}  & \frac{v_{\rm D}^4 \lambda}{2 Z \varphi_{\rm min}^2} + m_{\varphi}^2
    \end{pmatrix} \ ,
    \label{Eq:M_explicit}
\end{align}
where the physical dilaton mass
\begin{align}
m_{\varphi}^2 & \equiv \frac{1}{Z} \frac{\partial^2 V_{\rm eff}(\varphi)}{\partial \varphi^2} \bigg |_{\varphi_{\rm min}} = \frac{8 \pi^2}{3 N^2}  (1-n) \lambda_\varphi \varphi_{\rm min}^2 \ ,
\label{mphi_eqn}
\end{align}
as obtained from the stabilizing potential defined in the main text.

\begin{figure*}[ht]
    \centering
    \includegraphics[width=0.95\textwidth]{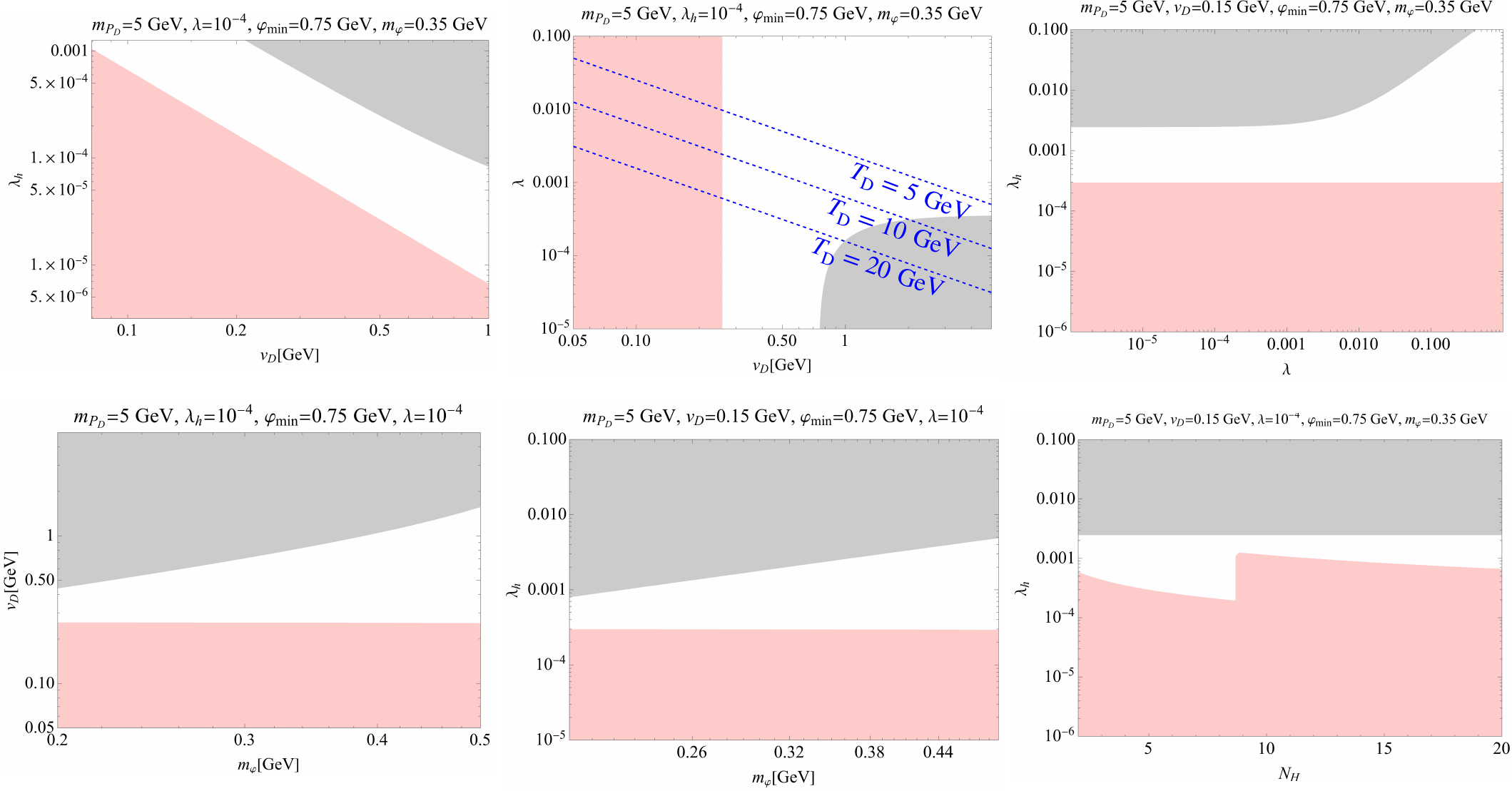}
    \caption{Direct detection exclusion limit (gray) and BBN constraint (red) on different parameter planes for illustrative choices of other parameters mentioned in the plot label. For the BBN constraint, other than the right bottom plot,
    we take $N_{\rm H}=5$.}
    \label{fig:para_dependence}
\end{figure*}

One can perform an orthogonal basis rotation to make the mass matrix diagonal, which defines the mass basis,
\begin{align}
    \Phi_i = O_{ij} \Psi_j \ ; \quad {\cal M}_{\rm diag} = O^{\rm T} \mathcal{M}  O = \begin{pmatrix}
        m_1^2 & & \\
        & m_2^2 & \\
        & & m_3^2 
    \end{pmatrix} \ .
    \label{Eq:basis_change}
\end{align}
The eigenvectors of ${\cal M}$ act as the columns in the orthogonal basis changing matrix $O$,
while the eigenvalues of ${\cal M}_{\rm diag}$ define the squared masses for the physical propagators.
Using this mass basis, we will calculate the relevant phenomenological amplitudes.

Let us start by calculating the BSM contribution to the SM Higgs invisible decay width.
Both $H \to \chi \psi$ and $H \to f \bar f$ contribute to this process,
\begin{align}
    \Gamma^{\rm BSM}_{h \to {\rm inv}} \simeq \left[ \left| \frac{m_\chi}{v_{\rm D}} O_{21} \right|^2 + \left| \frac{m_f}{\varphi_{\rm min}} O_{31} \right|^2 \right] \frac{m_h}{8 \pi} \ ,
    \label{Eq:GammaH}
\end{align}
where $m_f$ is the Dirac mass of $f$, expected to be at the cut-off $\sim \rm GeV$, $m_\chi$ is the mass of $\chi$,
and we have defined $\Psi_1$ to be the SM Higgs boson, which is mostly composed of the $h$ flavor state.
It is found that the current experimental constraint~\cite{ATLAS:2023tkt} implies $\lambda_h \lesssim 0.1$. 

We now discuss the direct detection of the DM candidate $p_{\rm D}$. As a result of the Higgs portal interaction, one effectively generates the following operator describing the interaction with any SM quark $Q$,
\begin{equation}
    {\cal L}_{{\rm eff}, p_{\rm D} Q} = C_{p_{\rm D} Q} (q^2) \  \overline{p}_{\rm D} p_{\rm D} \bar Q Q \ ,
\end{equation}
where $q$ is the momentum transfer in the process involved and the effective Wilson coefficient is
\begin{align}
    C_{p_{\rm D} Q} (q^2) & = \frac{m_{p_{\rm D}}}{\varphi_{\rm min}} \left( \sum_{i=1}^{3} \frac{O_{3i} O_{1i}}{q^2+m_i^2} \right) \frac{m_Q}{v} \ ,
    \label{Eq:CPDQ}
\end{align}
where $m_{Q}$ denotes the mass of the SM quark $Q$.
As a side note, this same operator can be used to model the collider detection prospect of this DM,
however, the DM direct detection constraint implies an upper bound on $\lambda_h$
which is much stronger than any collider constraint.
One matches the operator with Eq.~\eqref{Eq:CPDQ} at the nucleon level
to obtain the DM proton ($p$) or neutron ($n$) spin-independent effective scattering cross-section,
\begin{align}
        \sigma_{\rm SI}^{{\cal N}} = \frac{4 \mu_{p_{\rm D} {\cal N}}^2}{\pi A^2} \left[ Z f_p + (A-Z) f_n \right]^2 \ ,
    \label{sigma_SI_eqn}
\end{align}
where $\mu_{p_{\rm D} {\cal N}}$ is the DM-nucleon reduced mass, ${\cal N}$ represents both proton and neutron,
$A$, $Z$ are respectively the mass and atomic numbers of the target material
and the nuclear matrix elements are defined as
\begin{align}
    \nonumber
    \frac{f_{\cal N}}{m_{\cal N}}  & = \sum_{Q=u,d,s} \frac{C_{p_{\rm D} Q}}{m_Q} f_{T_Q}^{{\cal N}} + \frac{2}{27} f_{T_G}^{{\cal N}} \sum_{Q=c,t,b} \frac{C_{p_{\rm D} Q}}{m_Q} \ , \\[1ex]
    m_{\cal N}  f_{T_{Q}}^{\cal N} & \equiv \langle {\cal N} | m_Q \bar Q Q | {\cal N} \rangle \ ; \quad f_{T_G}^{\cal N} = 1- \sum_{Q=u,d,s} f_{T_Q}^{\cal N} \ .
    \label{fpfn_eqn}
\end{align}
The relevant momentum transfer for the direct detection of DM is $q \sim m_{p_{\rm D}} v_{\rm DM}$,
where $v_{\rm DM} \sim 10^{-3}$ is the local DM velocity.
The resultant constraints from the direct detection are shown in the main text,
which provide an upper limit on the portal coupling $\lambda_h$.

A lower bound on the portal coupling is inferred by the requirement that the massive dark pion $\pi_{\rm D}$,
produced as a result of the annihilation of the symmetric component of DM and anti-DM,
decays to SM final states before the onset of BBN,
which we take to be roughly 1s.
By naive dimensional analysis, the dilaton-dark pion coupling is expected to be
\begin{equation}
    {\cal L}_{\varphi \pi_{\rm D}} \approx \frac{\Lambda_{H,0}^2}{4 \pi} \varphi \pi_{\rm D} \ ,
    \label{Eq:piDphi}
\end{equation}
where $\Lambda_{H,0}$ denotes the confinement scale of the stabilizing dark $SU(N_{\rm H})$.
Hence, the decay rate of $\pi_{\rm D}$ is
\begin{align}
    \nonumber
    \Gamma_{\pi_{\rm D}} = \sum_{Q=u,d,c,s} & \left[ \frac{\Lambda_{H,0}^2}{4 \pi} \left(\sum_{i=1}^{3} \frac{O_{3i} O_{1i}}{m_{\pi_{\rm D}}^2-m_i^2}\right) \frac{m_Q}{v}\right]^2 \\
    & \times \frac{m_{\pi_{\rm D}}}{8 \pi} \left[1 - \frac{4 m_{Q}^2}{m_{\pi_{\rm D}}^2} \right]^{3/2} \Theta \left( m_{\pi_{\rm D}}-2 m_Q \right) \ ,
    \label{Eq:piD_decay}
    \end{align}
where $\Theta$ is the Heaviside step function.
We require $\Gamma_{\pi_{\rm D}}^{-1} \lesssim 1$s to be consistent with the BBN bound.
To relate the dark pion mass and the dark proton mass,
the following naive scaling expectations can be used~\cite{Nakai:2015ptz, Chivukula:1989qb}:
\begin{align}
    \nonumber
    m_{p_{\rm D}} & \simeq  m_p \frac{\Lambda_{H,0}}{\Lambda_{\rm QCD}} \frac{N_{\rm H}}{3} \ , \\[1ex]
    m_{\pi_{\rm D}} & \simeq  m_\pi \sqrt{\frac{2 m_f}{(m_u+m_d)} \frac{\Lambda_{H,0}}{\Lambda_{\rm QCD}}}  \ ,
    \label{Eq:mpdmpiD_naive}
\end{align}
where $m_p$, $m_\pi$ are the proton and pion mass, respectively, $\Lambda_{\rm QCD}$ is the QCD dynamical scale,
and $m_f$ is the Dirac mass of the constituent fermion $f$ in our model.
We use this relation to express $m_{\pi_{\rm D}}$ in terms of $m_{p_{\rm D}}$
and depict the BBN and DM constraints in the same plot in the main text.
Fig.~\ref{fig:para_dependence} shows the current direct detection exclusion limit in gray and the BBN constraint in red in different parameter planes to demonstrate the dependence on the parameter space. Note that $\chi, \psi$ can decay through the neutron portal before BBN as explained in the main text.
    
\bibliographystyle{apsrev4-2}
\bibliography{bib}